\documentstyle[preprint,aps,psfig]{revtex}

\begin{document}
\tightenlines

\title{Stretching Single Domain Proteins: Phase Diagram and Kinetics of
Force-Induced Unfolding}

\author{D. K. Klimov and D. Thirumalai}

\address{Institute for Physical Science and Technology and Department
of Chemistry and Biochemistry\\
University of Maryland, College Park, MD 20742}

\maketitle

\begin{abstract}

Single molecule force spectroscopy reveals unfolding of domains in titin
upon stretching. We provide a theoretical framework for these experiments
by computing
the phase diagrams for force-induced unfolding of single domain proteins
using lattice models. The results show that two-state folders (at zero force)
unravel cooperatively whereas stretching of non-two-state folders occurs
through intermediates.  The  stretching rates of
individual molecules show great variations reflecting the heterogeneity of
force-induced unfolding pathways. The approach to  the 
stretched state occurs in a step-wise "quantized" manner.
Unfolding dynamics depends sensitively on topology. 
The unfolding rates increase exponentially with force $f$ till an optimum value
which is determined by the barrier to unfolding when $f=0$. A
mapping of these results to proteins shows qualitative agreement with 
force-induced unfolding of Ig-like domains in titin. We show that
single molecule force spectroscopy can be used to map the folding free
energy landscape of proteins in the absence of denaturants.

\end{abstract}

\newpage

\section{Introduction} 

Titin, a giant protein molecule responsible for elasticity of muscles, is
comprised of a few hundred immunoglobulin (Ig) and fibronectin-III repeats
aligned in tandem \cite{Pan,Labeit,Erickson94}. 
Recently, through 
nanomanipulation of single protein molecules,  there has been direct evidence
for sequential unfolding of individual domains upon stretching 
\cite{Bustamante,Simmons97,Rief}. These remarkable experiments and others on
DNA \cite{Stretch94} have made it possible to unearth the microscopic 
underpinnings of the unusual elastic behavior in biological molecules. 
In two experiments \cite{Bustamante,Simmons97}
individual titin molecules were tethered to a
plastic bead and optical tweezers were used to stretch the molecule. Direct 
measurement of the forces required to stretch titin 
were used to infer that tension leads to unfolding of individual
Ig-like domains \cite{Bustamante,Simmons97}. Perhaps, 
the clearest evidence for domain
unraveling was presented by Rief {\em et al.} \cite{Rief}  
who used atomic force microscopy
(AFM) to pull on titin molecules adsorbed onto a gold surface. The AFM
experiments, on both the model recombinant titin molecules consisting 
only of Ig (Ig$_{4}$ and Ig$_{8}$) domains and the native titin, 
showed clear saw-tooth patterns in the
force-extension curves indicating sequential unfolding of domains. The
constant periodicity of the saw-tooth pattern ($\approx 25 nm$) is nearly
coincident with the dimensions of the fully unfolded Ig domain ($\approx 
29 nm$) and is very similar to the contour length inferred from fitting the
force-extension curves obtained from the optical tweezer experiments 
\cite{Bustamante,Simmons97} (see also related experiments on tenascin 
\cite{Erickson98}).
All the  experiments conclude that
sequential unraveling of the domains results upon mechanically 
stretching titin.

Inspired by these experiments, we report the results of 
force-induced unfolding of single domain proteins using simple
lattice models which have been useful in the search for general principles
of protein folding \cite{Dill97}. 
Because the primary mechanism of stretching titin
involves unraveling of individual Ig-like domains, which fold spontaneously
in the absence of tension \cite{Fong96,Goto}, our calculations 
provide  microscopic origins of force induced unfolding. We show that the
response of proteins to force depends primarily on their topology in the
absence of force.
By computing the phase diagram and kinetics of a
number of model proteins  subject to tension we show that 
the folding free energy landscape \cite{Dill97,Wolynes95} in 
the absence of
force can be deciphered using single molecule manipulation techniques.

\section{Materials and Methods }

The polypeptide chain is modeled as a sequence of $N$ connected beads
on a cubic lattice. The  energy of a  
conformation (given by the vectors ${\vec{r}_i}$ with $i =1,2,3,.....N$) is 
\begin{equation}
E_p=\sum _{i=1}^{N-3}\sum _{j=i+3}^{N} B_{ij}\delta(|\vec{r}_i-\vec{r}_j|-a)
\label{Ep}
\end{equation}
where $\delta(x)$ is the Kronecker delta function, $a$  
is the lattice spacing,  and $B_{ij}$ is the contact
energy between beads $i$ and $j$. The set of matrix
elements $B_{ij}$ specifies a sequence. We use two types  of contact
potentials, the statistical potentials derived by Kolinski, Godzik, and 
Skolnick (KGS) \cite{KGS93}  and the random bond (RB) 
model \cite{KlimThirum96}. 
The applied force 
to the terminal beads yields an additional energy  
\begin{equation}
E_f = -fz,  
\label{Ef}
\end{equation}
where $z =| \vec{r}_1 - \vec{r}_N |$ is the extension. Since the polypeptide
chain is on lattice, where continuous overall rotations are not possible, we
assume that upon stretching there is alignment of the protein along the
force direction with zero torque. This is equivalent to the assumption that
the relaxation time for the overall rotational degrees of freedom is much
shorter than that for structural relaxation that is responsible for
unfolding or folding processes. Thus we take the absolute
value of $z$ to represent the energy due to stretching. The total energy of
the chain conformation is given by the sum of Eqs. (\ref{Ep}) and (\ref{Ef}).
We use dimensionless units for energy whose typical value is in the 
range $(1-2)\,k_BT$; length is measured in units of $a$ (=0.38$nm$), 
and temperature is
measured in units of energy/$k_B$. For purposes of mapping these to physical 
values we use $2k_BT$ for energy which means force in our simulations
is measured in multiples of about $20\,pN$.

The thermodynamics as a function
of force ($f$) and temperature ($T$) is obtained using a variant of
the multiple histogram method
\cite{Ferrenberg} in conjunction with standard Metropolis Monte Carlo
simulations \cite{Metropolis}. 
When $f > 0$ the collection of histograms at different temperatures and zero
force  becomes unreliable
because highly stretched conformations are almost never sampled. Such
states, which have negligible Boltzmann weight in the absence of force, become
thermodynamically important upon stretching. It proves more convenient to
collect histograms at a fixed value of $T$ and at various values of $f$ so that
all relevant states across the entire $(f,T)$ plane are adequately sampled. 

To characterize the degree of similarity of an arbitrary sequence
conformation with the native structure we use the overlap function \cite{Cam93}
defined as   
\begin{equation}
\chi = 1 - \frac{1}{N^2-3N+2} \sum_{i\ne j,j\pm 1} \delta(r_{ij} - r_{ij}^0),
\end{equation}
where $r_{ij}$ is the distance between the beads $i$ and $j$, 
$r_{ij}^0$ is the corresponding distance in the native conformation,
and $\delta (x)$ is the Kronecker delta function. 

The kinetic simulations 
of force-induced unfolding were performed  at a constant
temperature  $T_s$ (below $T_F$, the folding temperature) that satisfies 
the condition $<\chi (f=0,T_s)> = 0.15$. 
Starting from the native conformation 
the force was suddenly increased to $f_s$ so that at 
$(f_s,T_s)$ stretched rod-like conformations are stable.
The unfolding kinetics is 
monitored by computing the distribution of stretch times, $\tau _{s,1i}$,
which is the first instance a
trajectory $i$ reaches a stretched state with no contacts. 
Typically $M=800$ trajectories have been 
generated for the calculation of unfolding rate. From the distribution
of stretch times the mean unfolding time and the fraction of folded molecules
at time $t$ can be calculated \cite{KlimThirum96}. These probes together with 
the dynamics of rupture of tertiary contacts and 
the time dependence of extension
are used to obtain the unfolding pathways. The mean unfolding time is
\( \tau_{u}  =  \frac{1}{M} \sum_{i=1}^{M} \tau_{s,1i} \)
the inverse of which is the unfolding rate $k_u$.

In order to obtain the general characteristics of force-induced unfolding we
computed the phase diagram and kinetics for five 
sequences (four 27-mers and one 36-mer) and differing 
interaction potentials. We chose four 27-mer and one 36-mer sequences, 
whose thermodynamic and kinetic characteristics in 
the absence of force 
are documented elsewhere \cite{KlimThirum96,KlimThirum98b},
to investigate unfolding
transitions upon stretching. Three of the 27-mer and 36-mer 
sequences fold kinetically and
thermodynamically by two-state mechanism. These sequences
have small values of $\sigma  = (T_{\theta} -T_{F})/T_{\theta }$, 
where $T_\theta $ and $T_{F}$ are the collapse and folding transition
temperatures respectively when $f = 0$ \cite{KlimThirum96,KlimThirum98b}. 
The fourth 27-mer sequence with $B_{ij}$ given by RB potentials 
has $\sigma  =0.11$, and its
folding (thermodynamics and kinetics) reveals intermediates
\cite{KlimThirum96}. Thus, with
these sequences, we can investigate the effect of stretching for
protein-like models that display distinct folding mechanisms in the absence
of force. For purposes of illustration we present results for a 36-mer sequence
with the KGS potentials ($\sigma \approx  0$) and the 27-mer RB model
sequence with $\sigma \approx  0.11$.

\section{Results}

The phase diagram for the 36-mer, which exhibits two-state cooperative thermal
unfolding when $f = 0$, is given in Fig. (1a). The states of the
polypeptide chain are represented by the thermal average overlap function 
$<\chi (f,T)>$, where $\chi $ gives the degree
of similarity to the native state. 
In particular, small values of the overlap
function correspond to conformations that belong to the native basin of 
attraction (NBA). The color codes
in Fig. (1) are such that red corresponds to small $<\chi (f,T )>$
(high native content and folded states), while the blue region has
large $<\chi (f,T)>$ representing unfolded states.
We see from Fig. (1a) that the  $(f,T)$ plane
divides into predominantly red (folded states) and blue
(unfolded states) regions. In the red region, the overlap $<\chi (f,T)>
\lesssim  0.1$ and the probability of being in the  
NBA is greater than 0.5
\cite{KlimThirum98b}. In the blue region $<\chi (f,T)>$ 
is typically greater than 0.8 and the probability of being in the NBA
is almost zero. 
There is only a narrow band of the green region which suggests that
the force-induced unfolding transition for this sequence is an all-or-none
process with no signature of intermediates. 

The sharp boundary between
folded and unfolded states resembles that of the type I superconductors in
the $(H,T)$ plane where $H$ is the applied magnetic field. 
With this analogy the
locus of points separating the NBA and the unfolded states is given by 
\begin{equation}
f_{c} \sim f_0\biggl(1 - \biggl(\frac {T}{T_{F}}\biggr)^\alpha \biggr),
\label{fc}
\end{equation}
where $f_{c}$ is the critical force required to unfold the protein, $f_0$
is the value of $f_{c}$ at $T = 0$, and $T_{F}$ is the folding transition
temperature at zero force. Both $f_0$ and $\alpha $ depend on the sequence
and the native state topology. 
The fit using Eq. (\ref{fc}) gives $f_0 \simeq  0.98$ and 
$\alpha \simeq  6.0$ for the 36-mer. An independent 
estimate for $f_0$ can be made
by using $|E_0| \approx  f_0\Delta L$, 
where $E_0$ is the energy of the native state and $\Delta L$ is the
gain in the end-to-end distance 
of the polypeptide chain upon stretching  to a fully extended rod 
state. For the 36-mer 
$E_0 = -30.4$ and  $\Delta L \approx  30.9$ that leads to $f_0\approx  0.98$.

The phase diagram for a sequence whose folding (in the absence of force)
involves intermediates is shown in Fig. (1b). Although the general
appearance is similar to that shown in Fig. (1a) there
are clear differences. The region of stability of the NBA (red region) is
confined to low temperatures and small forces. 
Secondly, the boundary 
between the folded and unfolded states is  fuzzy and contains a broad
green region. This suggests that the force (or temperature)-induced
unfolding is likely to be non-cooperative involving intermediates. This is
reflected in the force-extension curves  
which show signatures of intermediates (D. Klimov \& D. Thirumalai,
unpublished).  
In contrast, for two-state folders with  
$\sigma \approx  0$, the force-extension curves show that at 
$f \approx f_0$ the chain abruptly unfolds to a stretched conformation without
populating any detectable intermediates.  
In fact, the unfolding transition 
occurs in an extremely narrow interval of force which for
the 36-mer is $0.01f_0$ (D. Klimov \& D. Thirumalai,
unpublished). 

We shall point out that the
contact interaction energies (or more precisely the potential of mean
force) are dependent on temperature. It has been argued that the
temperature dependence of contact interactions has to be included in
order to reproduce certain experimental observations  in proteins
\cite{Chanbook}. However, we expect the qualitative features of the
phase diagrams seen in Fig. (1) will be observed experimentally
regardless of the details of the interaction energies.

How is the completely stretched conformation
kinetically reached starting from the native conformation when $f$
exceeds $f_{c}$? 
For the four sequences with small $\sigma $ (two-state folders in
the absence of force) we find that, when averaged over an ensemble of initial
molecules, unfolding occurs in a single kinetic step. However, 
there is a great variation in the time scales of stretching
to the rod-like state. This is
dramatically illustrated in Fig. (2a) in which we plot, for the 36-mer, 
extension $z(t)$   as a function of $t$
measured in Monte Carlo steps (MCS). There is a large unexpected
heterogeneity in the approach to the stretched state. A striking feature in
Fig. (2a) is that there is a large variability in the times 
taken to exhibit  significant stretching prior to reaching
the rod-like conformation. Global unraveling takes place 
cooperatively with the disruption of local and non-local contacts occurring
in an all-or-none manner. These features are masked in the ensemble average
$<z(t)>$ which is shown as a dashed line in Fig. (2a). There is a remarkable
similarity between the response of these protein-like models to
force and  that exhibited by flexible polymers subject to sudden elongational
flow \cite{Chu}. Another interesting feature of the unfolding dynamics is that,
just as in experiments, the force-extension curves show hysteresis during
the stretch-release cycles (D. Klimov \& D. Thirumalai, unpublished).

The time evolution of the distribution of extension values $z$ 
for 400 trajectories is plotted in Fig. (2b). This plot 
shows that on time scales less than mean stretch time the chain explores 
a diverse manifold of  states each with different $z$. 
Certain $z$ values have significantly larger probability
$P(z,t)$ than the others which suggests that unfolding occurs in a step-wise 
quantized
manner. Similar observations have been made using molecular
dynamics simulations of Ig-like domain \cite{Lu98}.

Despite the large variability in the stretch times $\tau _{s,1i}$ 
the mechanism of
approaching the rod-like conformation may be qualitatively described as
occurring in roughly three stages. On the time scale $\approx 
(0.1\,-\,0.5)\tau _{s,1i}$  
following the application of a sudden force there is a
loss of a number of native contacts. There is a concomitant increase in the
extension of the chain $z(t)/z_s \simeq  (0.1\,-\,0.5)$, where $z_s=N-1$. 
In the second stage, the sequence 
searches for the equilibrium rod-like conformation. There is great variation
in the time scale for this search. 
This stage is
characterized by one or several plateaus in $z(t)$ (see Fig. (2a)). Finally,
the chain explosively and cooperatively makes a transition to the rod
state with $z(t)/z_s \simeq  1$. Naturally, there are several 
exceptions to this
generic scenario. For example, curve (d) in Fig. (2a) shows that $z(t)$ reaches
its equilibrium value monotonically in an extremely short time. 

The dependence of the unfolding mechanisms on topology is illustrated   
by computing the dynamical evolution of all the topologically permissible
contacts. We describe the results for two
27-mers RB two-state folders labeled A and B  
(these are the sequences 61 and 63, respectively in \cite{KlimThirum96}). 
The native state of each sequence is  
maximally compact. However, the key topological distinction 
between them is that  in the native conformation for A 
the terminal beads are on the same facet of the cube, whereas for B they
are placed directly on opposite facets. 
By tracking the time evolution of the loss of the
156 topological contacts, of which 28 are native, we computed the breaking
times $\tau _b^k$ for contact $k$.  The times 
$\tau _b^k$ are determined by  $(i/N)(N-j)/N$ which
measures how close the contact $k$, formed between beads $i$ and $j$, is to the
sequence ends. For sequence A, we find that the time scales for the rupture
of contacts are similar for groups of contacts that are close to one or the
other end of the sequence. In contrast, for sequence B the disruption of
contacts from the amino terminus (bead 1) occurs fast, while the contacts
located near the carboxyl terminus break up later in the
unfolding process. Thus, topology 
determines details of the force-induced unfolding pathways.  
Since in Ig-like domain the amino and carboxy termini are at opposite ends
we expect that  
the underlying mechanism by which this domain unfolds may be similar to
that for sequence B.

In Fig. (3a) we present the force-induced unfolding rate 
$k_u$ as a function of $f$ for the 36-mer at $T_s = 0.49$.
Qualitatively similar results were obtained for the 27-mers as well. The 
unfolding rates were computed from the distribution of stretch times for
800 trajectories. It is 
expected that $k_u$ should increase upon increasing $f$ because the activation
free energy is lowered upon application of force \cite{Bell78,Evans97}. 
The free energy profiles as a function of the number of native contacts
($Q$), which is an approximate reaction coordinate for two-state folders
\cite{Socci96}, are given in Fig. (3b) for the 36-mer at various force values.
The decrease in the unfolding barrier explains the observed dependence of
$k_u$ on $f$ for $f < f_{opt}$. 
The unfolding rate is well 
described by $k_u \simeq  k_0\exp(f\Delta x/k_BT)$ for $f <  f_{opt}$
which is given by 
\begin{equation}
f _{opt} \sim \Delta F^{\ddagger }(T_s,f=0)/\Delta x, 
\label{fopt}
\end{equation}
where $\Delta F^{\ddagger }(T_s,f=0)$ is the unfolding free energy 
barrier at  $T = T_s$ and zero force, 
$\Delta x$ is an approximate width of the unfolding potential,  
and $k_{0}(T_s)$ is the unfolding rate in the absence
of force. For the 
36-mer $\Delta F^{\ddagger }$ is 2.26 at $T_s = 0.49$ (see Fig. (3b)), 
$f_{opt} \approx  3.2$ and therefore $\Delta x \approx  0.02L$ where $L$ is the
contour length of the chain. The small value of $\Delta x$ implies that the
transition region is quite narrow. When $f \geq 
f_{opt}$ the unfolding rate starts to decrease because sudden
(corresponding to large pulling speed) application
of relatively large forces traps the polypeptide chain in 
conformations, whose unfolding requires transient shortening of the
end-to-end distance. The transition from such conformations, which requires 
local annealing of the chain, slows down the
unfolding process. For $f >  7$ (see Fig. (3a)) there is a free
energy barrier associated with the breakup of contacts in the
conformations acting as  transient kinetic traps. In this force regime  
the fraction of molecules that are  folded at a time $t$ 
is best fit using a sum of two exponentials with the slow phase
signaling the onset of local trapping.

Our results for unfolding triggered by force are consistent with a
number of experimental observations on the unraveling of isolated Ig-like
domains in titin \cite{Bustamante,Simmons97,Rief,Erickson98}. 
(1) The ratio of $f_{c}/f_{0}$ for Ig-like domains
can be computed using Eq. (\ref{fc}) with $T = 25^{\circ }C$, 
the folding temperature 
$T_{F} \approx 60^{\circ }C$ \cite{Politou94}, and $\alpha \approx  6.0$. 
This gives 
$f_{c}/f_0 \approx  0.49$. From the phase diagram in Fig. (1a) we obtain
a similar value for the 36-mer when the temperature (measured in
Kelvin) is approximately $0.89T
_{F}$. Thus the general shape of the phase boundary should  be useful
in calibrating the experimental measurements on proteins. (2) The typical
values of the threshold force $f_0$ 
required to induce stretching in the two-state
folders are in the range of $(1\,-\,2.5)$. By translating these
into physical units we obtain 
$f_{0} \approx  (20\,-\,50)\,pN$. Using this  we
would predict that the unfolding force $f_c$, which depends on the pulling
speed, for Ig-like domains in titin should be around $(10\,-\,25)\,pN$. 
These  values are not inconsistent with the
experimental measurements (see Fig. (5) of \cite{Rief}). 
(3) The width of the unfolding potential $\Delta x$ which is obtained
using Eq. (\ref{fopt}) 
and the computed values of $f_{opt}$ and the activation free
energy $\Delta F^{\ddagger }$ (see Fig. (3b)) in the absence of force for the
36-mer is $0.02L$. Rief {\em et al.} \cite{Rief}
estimated that $\Delta x/L \approx  0.01$
using $\Delta x = 0.3\,nm$  and $L = 31\,nm$. 
If we use the fact that the lattice constant $a$, 
which gives the distance between $\alpha $-carbon atoms, is 
$\approx 0.4\,nm$ then the
value of $\Delta x$ for the 36-mer in physical units is roughly 
$0.3\,nm$. These
numbers are in very good accord with the experiments. 

The theoretical findings can be used to map the underlying
folding free energy landscape for two-state proteins using data from 
force-induced unfolding experiments. This is illustrated by applying
our results for the 36-mer to Ig-like domains. An estimate of $f_{opt}$ for
Ig domain can be made using the value of $3.2$ for the 36-mer, and by assuming 
that a $f_{opt}$ scales linearly with $N$. For the 90 residue Ig-like domain we
find that $f_{opt} \sim 160\,pN$. The unfolding barrier is $f_{opt}\Delta x$
which is approximately $12\,k_{B}T$ 
assuming that $\Delta x = 0.3\,nm$ \cite{Rief}.
From the stability of Ig domain ($\Delta G \approx 2.6\,kcal/mol$
\cite{Fong96}) 
we predict that the refolding barrier is approximately 
$4.6\,kcal/mol$. From these the folding
and unfolding times are predicted to be $0.1\,s$ and $400\,s$ 
respectively.  These
predictions are in fairly reasonable agreement with experimental estimates 
\cite{Fong96}. The estimates of barriers to folding by force-induced
unfolding measurements are likely to be complement 
to the standard method of
measuring rates at finite denaturant concentration and then extrapolating to
the desired values in the absence of denaturants.

\section{Conclusions}

This study has led to the following predictions: (i) The 
unfolding time scales should decrease
exponentially with force only till an optimum value of force, whose
magnitude is determined by only the unfolding barrier in 
the absence of force. 
(ii) The phase diagram, especially the boundary separating the unfolded and
folded states, has the characteristic type seen in  Fig. (1) and is 
quantitatively given by Eq.  (\ref{fc}).
(iii) The nature of
force-induced unfolding depends on the proximity of the amino and carboxy
termini in the native state. 
These predictions are all amenable to experimental test.
Because the response to force depends sensitively on the
characteristics of the sequence when $f$ is zero, it follows that
the mechanisms of protein folding (presence of
intermediates, the nature of the transition states,  and barriers to
folding) may be very directly probed by single
molecule force spectroscopy.

\acknowledgments
We are grateful to Harmen Bussemaker for discussions
during the early stages of this work. 
This work was supported by a grant
from the National Science Foundation through grant no. CHE96-29845.

\begin{figure}

\noindent
{\bf Fig. 1.} Phase diagrams in the $(f,T)$ plane  for (a) the 36-mer 
two-state KGS sequence   
and (b) 27-mer moderate folding RB sequence. The color
code for $<\chi(f,T)>$ is given on the right. The red color
corresponds to the states in the NBA, 
whereas blue color indicates the unfolded states. 
Green areas correspond to intermediate partially folded states. 
For both the sequences the boundary of NBA
is given by Eq. (\ref{fc}).

\noindent
{\bf Fig. 2.} (a) The time dependence of extension $z(t)/z_s$ for 201
individual trajectories of 36-mer sequence at $T_s=0.49$ and
$f_s=4.0$. Three generic trajectories (curves
(a)-(c)), the fastest  (d), and 
the slowest (e) trajectories are shown in black. The rest are given in
grey. The average $<z(t)>$ obtained from  
400 individual trajectories is represented by a thick
dashed curve.  \\
(b) The probability distribution $P(z/z_s,t)$ for 36-mer sequence under
the same unfolding conditions as in Fig. (2a). $P(z/z_s,t)$ gives the kinetic
probability of occurrence of the extension value $z$ at the time $t$. 
In both plots the extension values are normalized by
$z_s=N-1$. 

\noindent
{\bf Fig. 3.} (a) The dependence of unfolding rate $k_u$ (filled circles)
on $f$ for the 36-mer sequence at $T_s=0.49$. 
For  values of $f$ less than 7.0 
the force-induced unfolding takes place by a two-state process.  
For $f > 7.0$ unfolding trajectories separate
into a fast pool (the fraction of which is $\Phi $) that reaches 
the stretch state with the rate  $1/\tau _{s,FAST}$ 
(open circles) and a slow pool trajectories characterized by the  rate 
$1/\tau _{s,SLOW}$ (open squares). At these $f$ the  partition
factor $\Phi < 1$.  \\
(b) The free energy $F(Q)/T_s$ for 36-mer sequence at
$T_s=0.49$ and $f=0.0$ (diamonds), $f=0.3$ (full
circles), $f=0.54$ (open circles), and $f=0.7$ (open squares). It is
seen that the free energy unfolding barrier $\Delta F^{\ddagger }/T_s$
decreases as $f$ becomes stronger:  $\Delta F^{\ddagger }/T_s = 4.6$ at
$f=0.0$, 4.0 at $f=0.3$, 3.3 at $f=0.54$, and 2.8 at $f=0.7$. The 
free energy of stability (in units of $T_s$) at $f=0.0$ is 2.0.

\end{figure}

\newpage

\begin{center}
\begin{minipage}{15cm}
\[
\psfig{figure=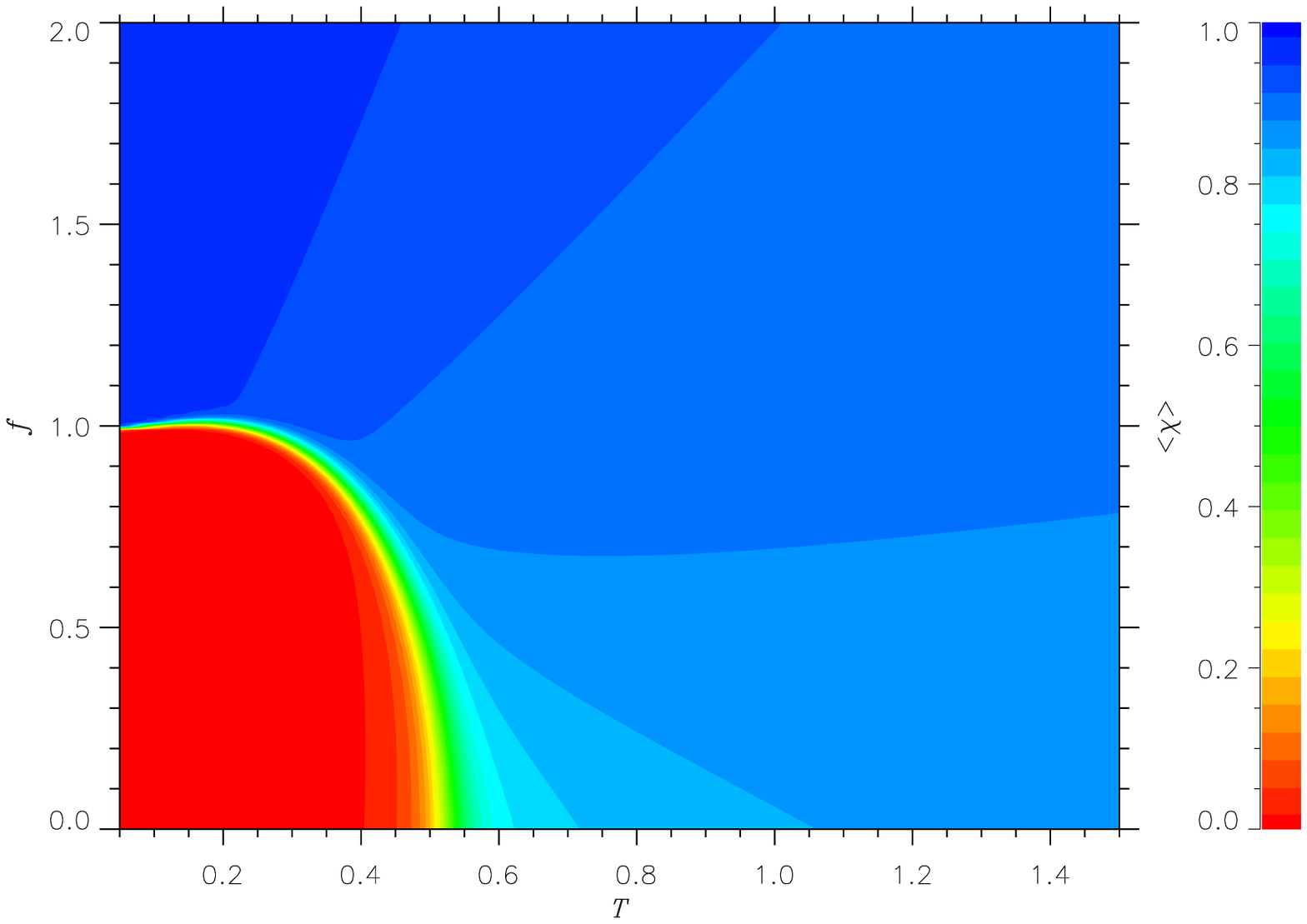,height=10cm,width=14cm}
\]
\end{minipage}

{\bf Fig. 1a}
\end{center}

\begin{center}
\begin{minipage}{15cm}
\[
\psfig{figure=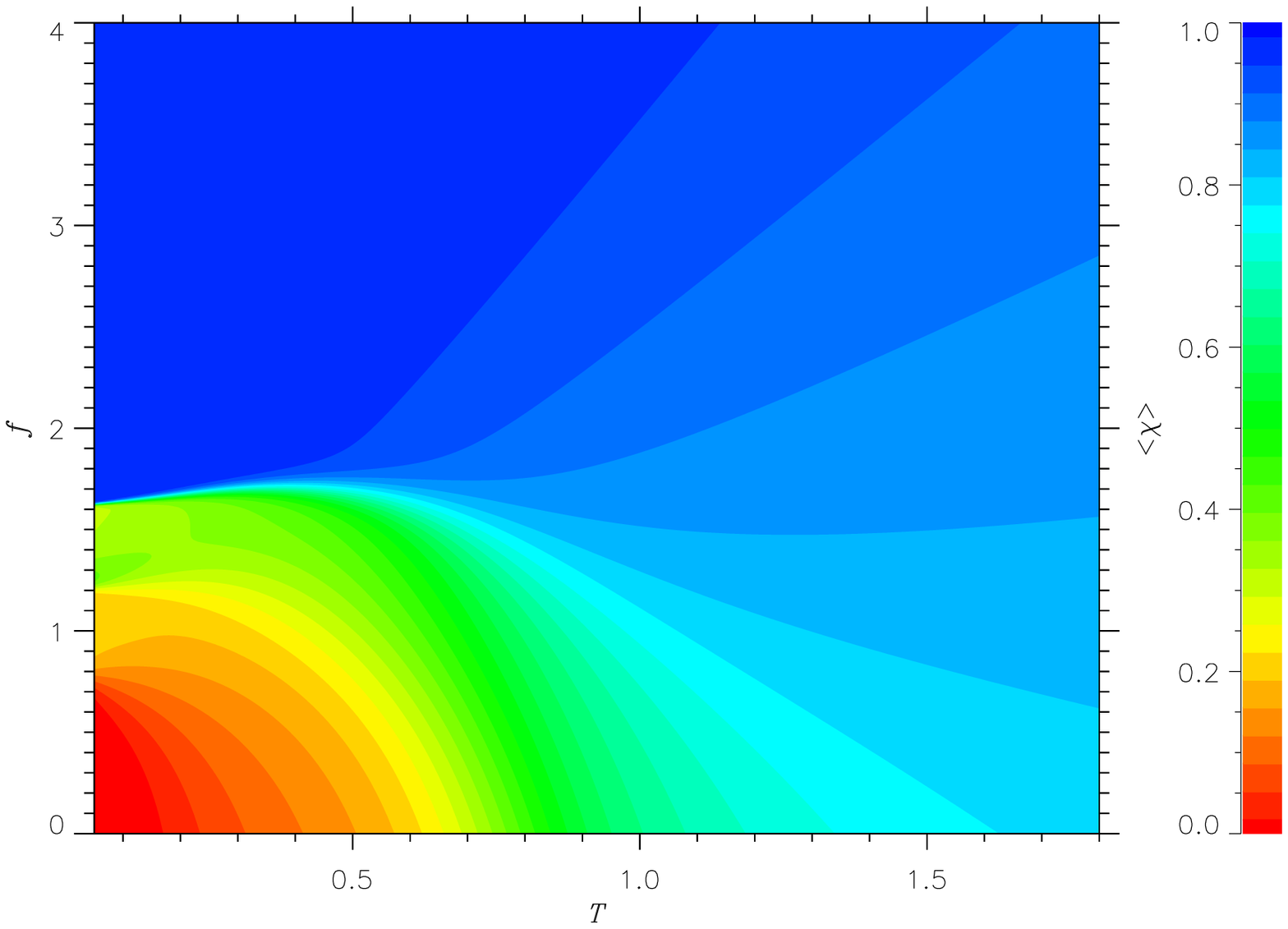,height=10cm,width=14cm}
\]
\end{minipage}

{\bf Fig. 1b}
\end{center}

\newpage

\begin{center}
\begin{minipage}{15cm}
\[
\psfig{figure=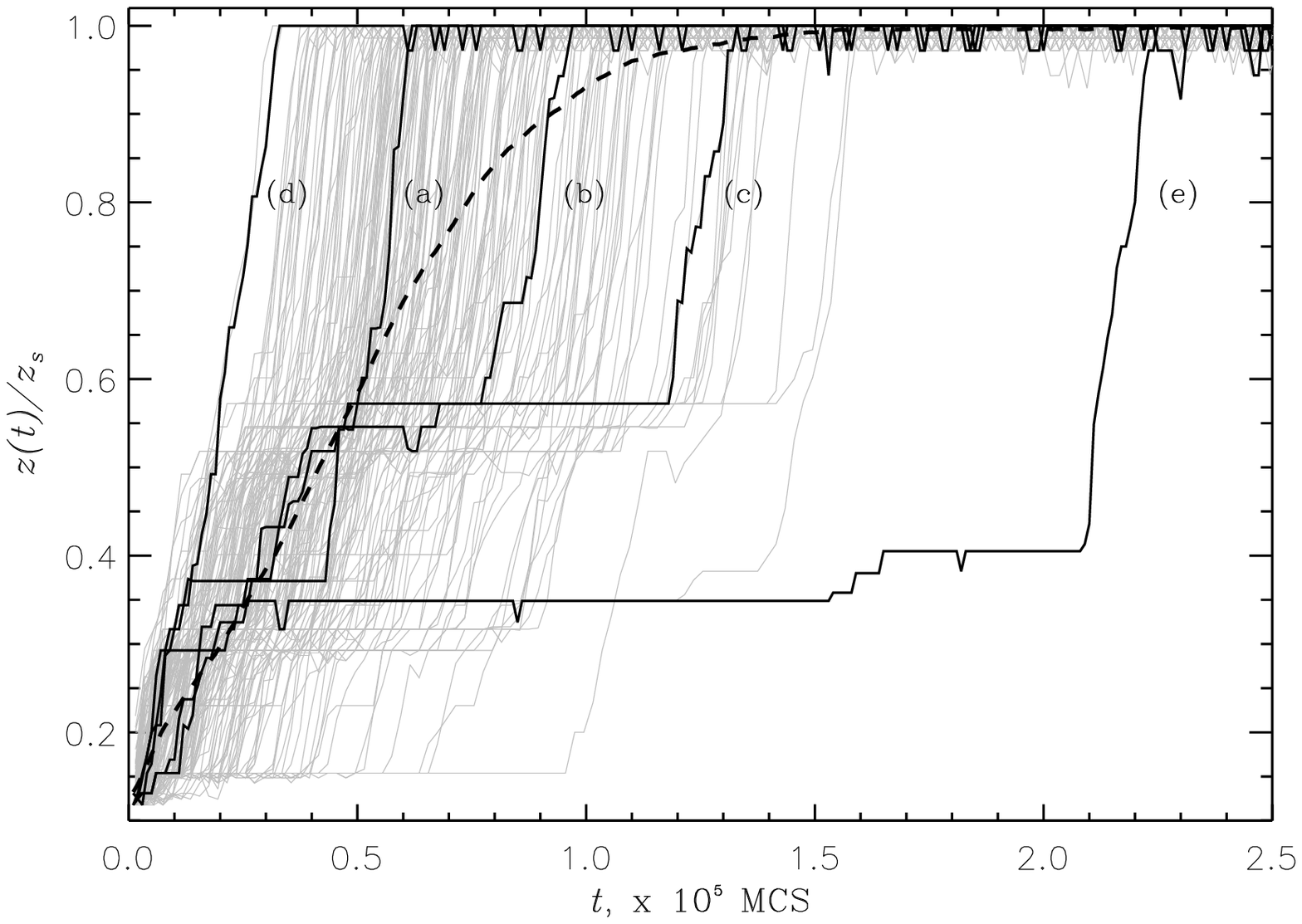,height=9cm,width=12cm}
\]
\end{minipage}

{\bf Fig. 2a} 
\end{center}

\begin{center}
\begin{minipage}{15cm}
\[
\psfig{figure=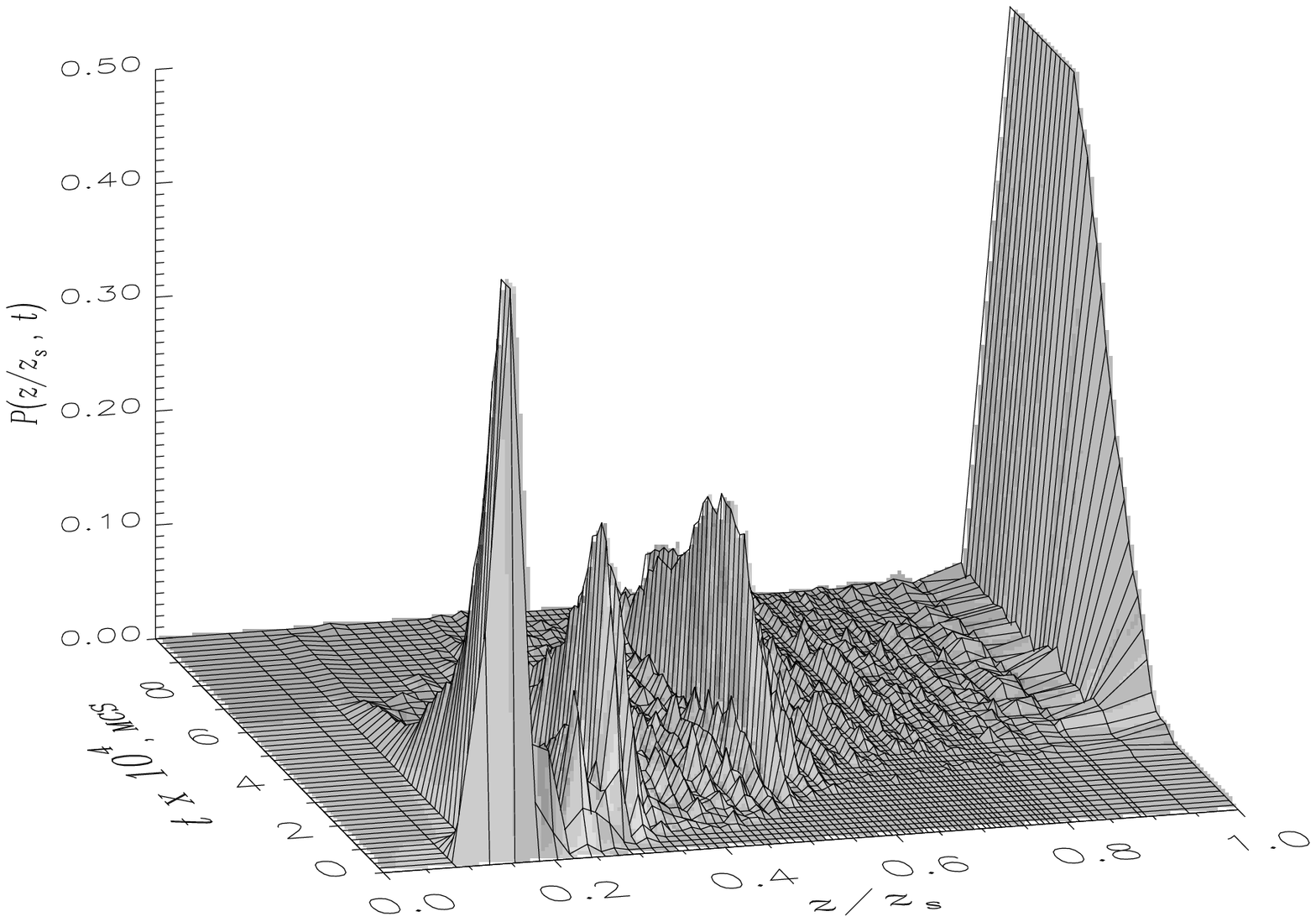,height=9cm,width=12cm}
\]
\end{minipage}

{\bf Fig. 2b} 
\end{center}

\newpage

\begin{center}
\begin{minipage}{15cm}
\[
\psfig{figure=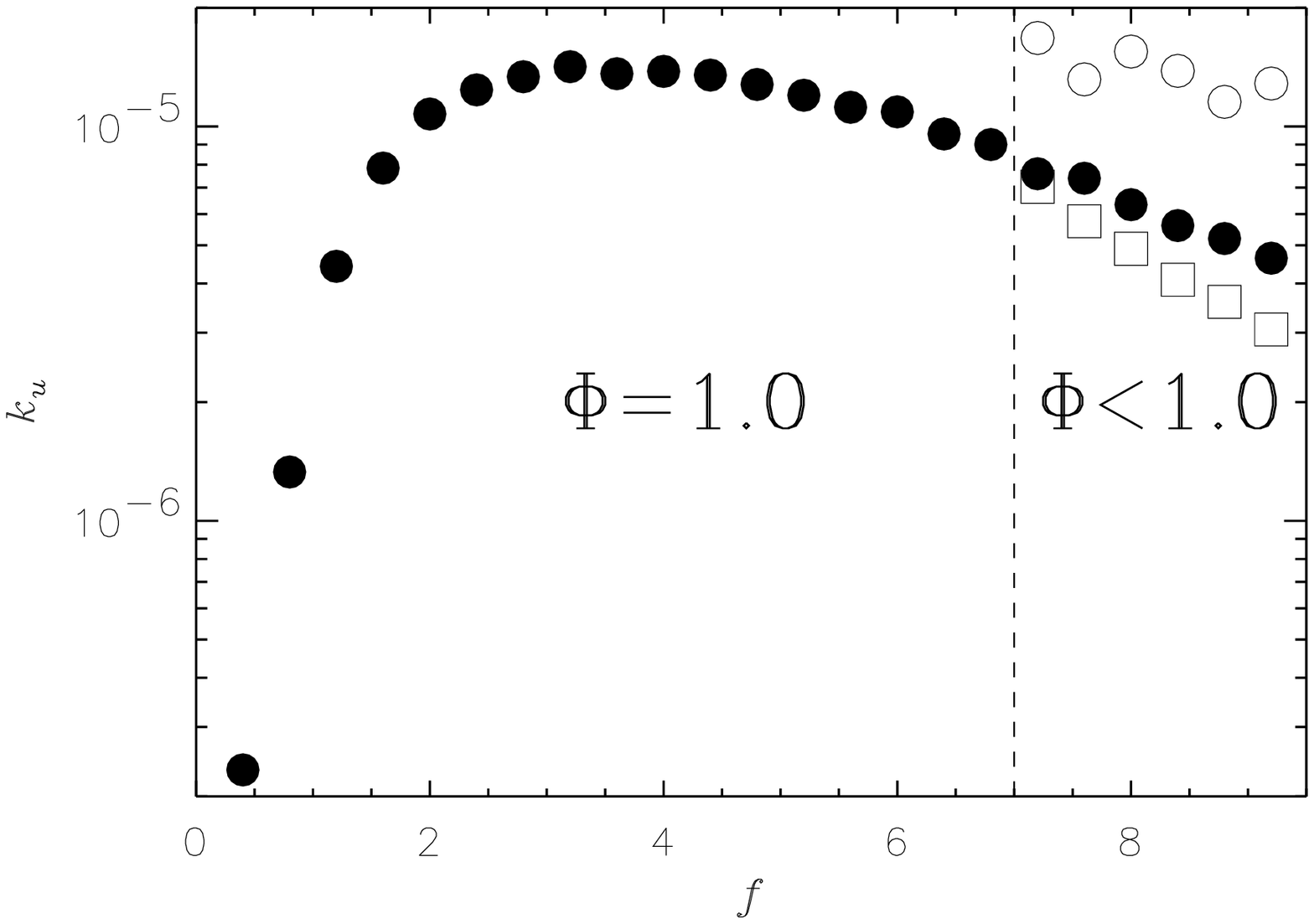,height=9cm,width=12cm}
\]
\end{minipage}

{\bf Fig. 3a} 
\end{center}

\begin{center}
\begin{minipage}{15cm}
\[
\psfig{figure=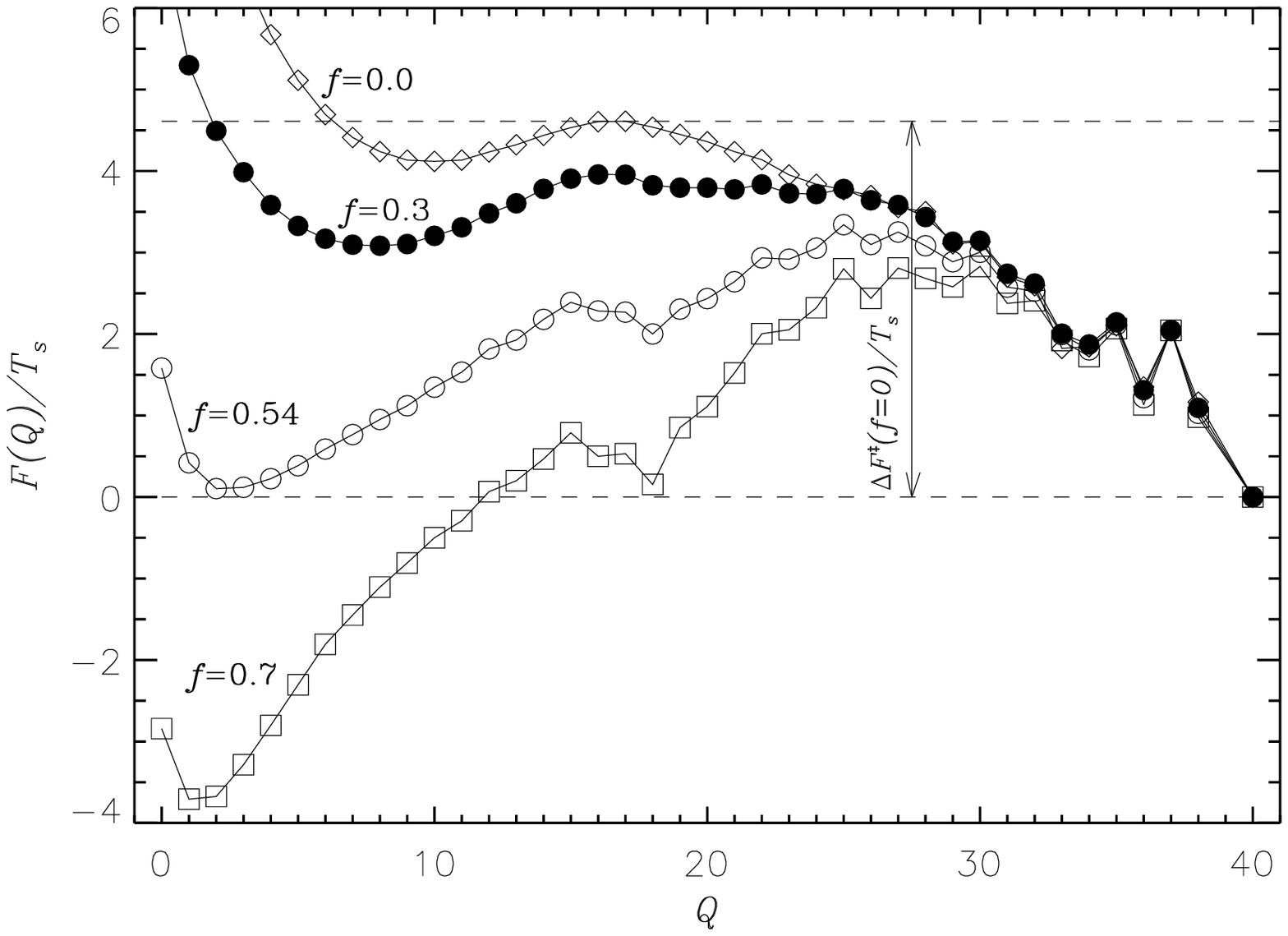,height=9cm,width=12cm}
\]
\end{minipage}

{\bf Fig. 3b} 
\end{center}

\end{document}